\documentclass[english]{article}

\usepackage{graphicx}
\usepackage{natbib}
\usepackage[normalem]{ulem}
\usepackage{amsfonts}
\usepackage{amsmath}
\usepackage[ruled,vlined]{algorithm2e}
\usepackage{thmtools}
\usepackage{thm-restate}
\usepackage{multicol}
\usepackage{amsthm}
\usepackage{mathtools}
\usepackage{amsbsy}
\usepackage{arydshln}
\usepackage{color}
\usepackage[T1]{fontenc}
\usepackage[utf8]{inputenc}
\usepackage[font=small,labelfont=bf]{caption}
\usepackage{multirow}
\usepackage[a4paper, total={6in, 8in}]{geometry}
\usepackage{authblk}
\usepackage{hyperref}
\usepackage{cleveref}

\newcommand{\var}{\operatorname{Var}}
\newcommand{\cov}{\operatorname{Cov}}

\newcommand{\E}{{\mathbb E}}

\makeatletter
\newcommand{\printfnsymbol}[1]{%
  \textsuperscript{\@fnsymbol{#1}}%
}
\makeatother

\begin{document}
	\bibliographystyle{agsm}
	\title{Monitoring the risk of a tailings dam collapse through spectral
analysis of satellite InSAR time-series data}
	
    \author[1*\printfnsymbol{2}]{Sourav Das}
    \author[2\printfnsymbol{2}]{Anuradha Priyadarshana}
    \author[3\printfnsymbol{2}]{Stephen Grebby}

    \affil[1*]{College of Science and Engineering, James Cook University, Cairns, QLD, Australia.}
    \affil[2]{Department of Statistics, Faculty of Applied Sciences, University of Sri Jayewardenepura, Nugegoda, Sri Lanka.}
    \affil[3]{Nottingham Geospatial Institute, Faculty of Engineering, University of Nottingham, Nottingham, NG7 2TU, U.K.}

    \date{}

    \maketitle

    \vspace{-1cm}
    \begin{center}
        \noindent\textsuperscript{*}Corresponding author(s). E-mail(s): \href{mailto:sourav2.das@googlemail.com}{\textcolor{blue}{\underline{sourav2.das@googlemail.com}}}
        \\Contributing authors: \href{mailto:anupriyadarsh@gmail.com}{\textcolor{blue}{\underline{anupriyadarsh@gmail.com}}};
        \href{mailto:stephen.grebby@nottingham.ac.uk}{\textcolor{blue}{\underline{stephen.grebby@nottingham.ac.uk}}};\\
        \noindent\textsuperscript{†}These authors contributed equally to this work. 
    \end{center}
        \vspace{0.5cm}
        
	\begin{abstract}
		Slope failures possess destructive power that can cause significant damage to both life and infrastructure. Monitoring slopes prone to instabilities is therefore critical in mitigating the risk posed by their failure. The purpose of slope monitoring is to detect precursory signs of stability issues, such as changes in the rate of displacement with which a slope is deforming. This information can then be used to predict the timing or probability of an imminent failure in order to provide an early warning. Most approaches to predicting slope failures, such as the inverse velocity method, focus on predicting the timing of a potential failure. However, such approaches are empirical and require some subjective analysis of displacement monitoring data to generate reliable timing predictions. In this study, a more objective, statistical-learning algorithm is proposed to detect and characterise the risk of a slope failure, based on spectral analysis of serially correlated displacement time-series data. The algorithm is applied to satellite-based interferometric synthetic radar (InSAR) displacement time-series data to retrospectively analyse the risk of the 2019 Brumadinho tailings dam collapse in Brazil. Two potential risk milestones are identified and signs of a definitive but emergent risk (27 February 2018--26 August 2018) and imminent risk of collapse of the tailings dam (27 June 2018--24 December 2018) are detected by the algorithm. Importantly, this precursory indication of risk of failure is detected as early as at least five months prior to the dam collapse on 25 January 2019. The results of this study demonstrate that the combination of spectral methods and second order statistical properties of InSAR displacement time-series data can reveal signs of a transition into an unstable deformation regime, and that this algorithm can provide sufficient early-warning that could help mitigate catastrophic slope failures.
		
	\end{abstract}
	
	%%%%%%%%%%%%%%%%%%%%%%%%%%%%%%%%%%%%%%%%%%
	\section{Introduction}
	Slope failures in the form of landslides or the collapse of engineered structures (e.g., tailings dams) pose a considerable risk to both life and infrastructure. Mitigating the risk they pose through providing an early warning relies critically upon monitoring slopes for precursory signs of instability \citep{intrieri2013brief}. Conventionally, this has involved using survey monuments, inclinometers, piezometers, extensometers and ground-based radar to monitor how a rock mass is deforming, with changes in the displacement and velocity providing the most reliable indication of the stability of a slope \citep{carla2017monitoring}. Accordingly, several empirical approaches have been developed to predict the timing of a slope failure based on displacement monitoring data \citep{intrieri2016landslide}. 
	
	The majority of failure timing prediction approaches utilise the concept of accelerating (tertiary) creep theory \citep{saito1969geomagnetic, fukuzono1985method}, under which the behaviour of a material in the terminal stages of failure under constant stress and temperature conditions is governed by the empirical power-law \citep{voight1988method, voight1989relation}:
	
		\begin{equation}\label{eq:failaw}
		v(t)^{-\alpha} \frac{d v(t)}{dt} = A 
	\end{equation}
	
	where $v$ is the creep velocity, $t$ is the time, and $\alpha$ and $A$ are constants. It has been observed that for a wide range of slope failures $\alpha \approx 2$ \citep{voight1989relation}. Consequently, for $\alpha$ = 2, equation \eqref{eq:failaw} has the solution: 
	
	\begin{equation}\label{eq:faillaw1}
		1/v(t) = A(t_f -t)
	\end{equation}  
	
	where $t_f$ is the time of failure. Equation \eqref{eq:faillaw1} implies that approaching the time of failure, the inverse of the velocity scales as a linear function of time. Consequently, the intercept point on the time axis of an inverse velocity (1/$v(t)$) vs. time ($t$) plot corresponds to the time of failure ($t_f$). This approach, often referred to as the inverse velocity method, is commonly applied to surface displacement monitoring data to predict the time of slope failures, by using linear regression to extrapolate the inverse velocity trend to the point of intersection on the time axis \citep{carla2018integration}.

	Although predicting the timing of slope failures can be crucial in mitigating the risk they pose, there are several factors that affect the ability to do so effectively. Firstly, the conventional ground-based monitoring techniques often provide measurements that are low density, have limited coverage and have a low or irregular temporal sampling frequency, which can make it difficult to detect the precursory tertiary creep \citep{carla2019perspectives}. Secondly, empirical prediction approaches like the inverse velocity method require expert user-intervention in order to derive reliable estimates of the slope failure timing. This includes filtering the velocity time-series measurements to remove instrumental and random "noise" to help determine the onset of tertiary creep (i.e., Onset Of Acceleration) and increase the fit of the linear regression line on the inverse velocity-time plot \citep{dick2015development, carla2017guidelines}. However, in the absence of a general rule regarding filtering, the selection of the optimal method is subjective and typically determined through trial-and-error \citep{rose2007forecasting}. Furthermore, the inverse velocity trend often only approaches linearity in the final few weeks prior to a slope failure \citep{rose2007forecasting}, therefore only enabling reliable short-term predictions that leave little time for risk mitigation measures to be implemented. 
	
	The use of satellite Interferometric Synthetic Aperture Radar (InSAR) has been increasingly recognised as an effective solution in overcoming the issue of limited surface displacement measurement coverage. The InSAR method is based upon the concept of interferometry, with precise changes in the surface elevation being derived from the interference of electromagnetic waves from two Synthetic Aperture Radar (SAR) acquisitions \citep{rosen2000synthetic}. Accordingly, satellite InSAR has been used to measured ground motion for a wide range of applications, including volcanology \citep{massonnet1995discrimination, pinel2014volcanology}, landslides \citep{fruneau1996observation,colesanti2006investigating}, seismology \citep{massonnet1993displacement, peltzer1995surface} and various other types of geohazard monitoring \citep{gee2019national}. More recently, satellite InSAR has also been applied to slope failure prediction in relation to landslides, tailings dams and open-pit mines \citep{intrieri2018maoxian,carla2019perspectives}. A particularly pertinent example concerns the 2019 catastrophic collapse of tailings Dam I in Brumadinho, Brazil. The dam collapsed causing a catastrophic mudflow that resulted in the death of more than 250 people, without any apparent risk indicators from the array of conventional methods being used to monitor its stability \citep{robertson2019report}. However, several studies have retrospectively analysed the stability of the dam preceding the collapse using satellite InSAR \citep{du2020risk,holden2020brumadinho}, with some of these deriving inverse velocity-based failure timing predictions from the precursory deformation \citep{f2020deformations, grebby2021advanced}. While successful predictions close to the actual time of failure were possible using the inverse velocity method, there remains some degree of uncertainty over the reliability of these predictions, due largely to the fitting of the linear regression line and the amount of prior warning of a potential failure that can be provided. 
	
	There have been attempts to generalise the inverse velocity method in equation \eqref{eq:failaw} by extending the temporal dynamics of the governing mechanisms of failure to consider locally (spatially) varying dynamics. Recent examples include studies by \cite{tordesillas2018data} and \cite{niu2021forecast} and references therein. Furthermore, there is a growing body of literature exploring the potential to offer probabilistic predictions of slope failure by modifying the inverse velocity method in various ways, such as through the application of linear regression \citep{zhang2020probabilistic}, stochastic differential equations \citep{bevilacqua2019probabilistic}, quantile-regression \citep{das2019near, wang2020modeling}, and mixture-distribution \citep{zhou2020early}. However, with the exception of \cite{bevilacqua2019probabilistic}, none of these studies explicitly consider the temporal serial correlation of the displacement or velocity fields. Additionally, almost all of these studies assume a parametric statistical distribution on the observed response (i.e., displacement or velocity). Nevertheless, rapid developments in technology for monitoring ground displacements, such as ground-based \citep{casagli2010monitoring} and satellite-based InSAR \citep{WASOWSKI2014103} present the opportunity to develop much faster and sensitive precursory warnings using algorithms that are based on streaming data and require less stringent statistical assumptions about the displacement mechanism. \cite{das2019near} proposed a data-driven, non-parametric multivariate statistical methodology for near real-time monitoring of slope failures. However, \cite{das2019near} acknowledged that serial correlation inherent in such deformation signals was not modelled. Nonetheless, their framework led to a much earlier prediction of the failure than that offered by the inverse velocity method. 
	
	Accordingly, the aim of this study is to further develop the framework of \cite{das2019near} to propose a new generalised algorithm that can accommodate temporal correlation whilst retaining its non-parametric nature. One key advantage of this statistical approach is that often regular sampling may reveal a change in the slope stability regime much earlier than that observed using the mechanistic failure law. The failure dynamics obtained by the algorithm is driven entirely by properties of the data and a fundamental statistical property underpinning conventional theory of time-series analysis called second order stationarity. Consequently, the statistical nature of this approach also makes it more objective in comparison to the inverse velocity method because the subjectivity associated with manually identifying the Onset of Acceleration and optimally filtering the time-series data is circumvented. 
	
	Heuristically, a second order stationary time-series has time-invariant statistical properties of second order. It is therefore postulated that the displacement time-series of a slope, sampled during a stable epoch, must be second order stationary. However, gradual deviations from second order stationary should occur as the slope transitions into an unstable epoch. Here, it is this principle that is utilised in order develop the new algorithm, which can be broadly defined as two phases:
	\begin{enumerate}
		\item Phase~1: Detecting a change in regime -- this involves first characterising (estimating) the second order statistical features of an area during a stable regime based on regularly sampled InSAR displacement time-series data. This is achieved through estimating statistical features of multiple time-series in terms of their dynamic periodograms. The stable regime (stationarity) is then defined as the entire history of the observations up until the detection of statistical variation in periodogram features; this point marks the time of regime change.
		\item Phase~2: Risk classification -- this involves continuously tracking the deviation of the time-series from second order stationarity after the time of regime change detect in Phase~1. Using a combination of supervised and unsupervised learning algorithms, this deviation is described as a concave probabilistic function of time. Risk of failure warnings are then generated based on estimates of inflection points on the trajectory of this function.  
	\end{enumerate}
	In this study, the proposed algorithm is applied to satellite InSAR time-series data to retrospectively analyse the risk of the Dam I tailings dam collapse. Section~\ref{sec:sec2} describes the study area and the InSAR time-series data used in the analysis. The proposed algorithm is then presented in Section~\ref{sec:mth}, before the results and discussion for its application are provide in Section~\ref{sec:res}. Finally, the key conclusions, limitations and future research associated with this study are outlined in Section~\ref{sec:conclusion}.
	
	%%%%%%%%%%%%%%%%%%%%%%%%%%%%%%%%%%%%%%%%%%
	
		\begin{figure}[!h]
		\centering
		\includegraphics[scale = 0.45]{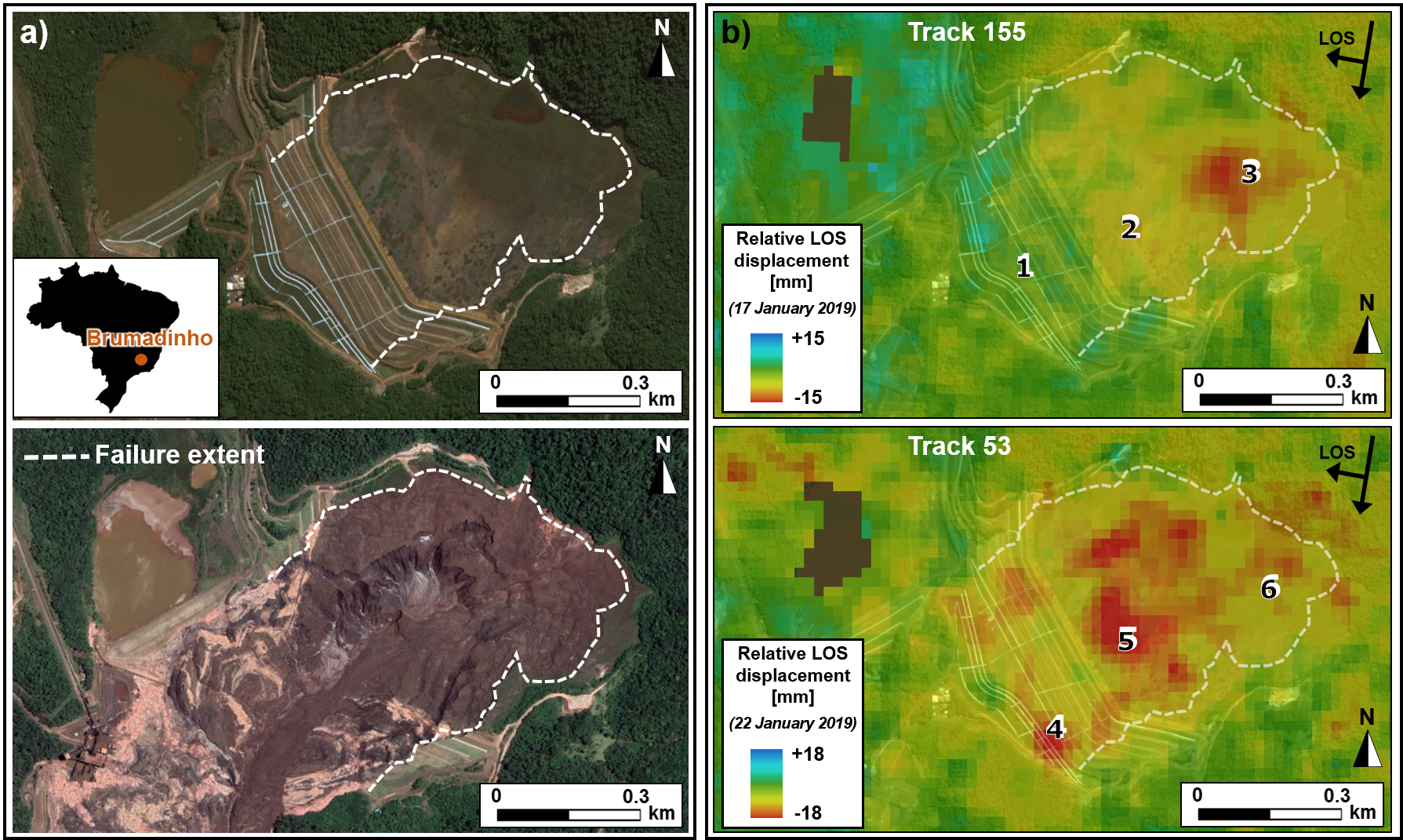}
		\caption{The Brumadinho (Brazil) study area. \textbf{a}) Satellite image of Dam I pre- and post-collapse, \textbf{b}) Relative LOS displacement maps for tracks 155 and 53 prior to the collapse, with locations 1-6 indicated.}
		\label{fig:studyA}
	\end{figure}
	
	\section{Study Area and Data}\label{sec:sec2}
	Dam I (Figure~\ref{fig:studyA}a), at the Córrego do Feijão iron ore mine complex in Brumadinho, Minas Gerais State (southeastern Brazil), was an inactive (since 2015) upstream tailings dam that underwent a sudden and complete failure across its full 720 m width and 86 m height on 25 January 2019 as a result of a flow (static) liquefaction mechanism. The expert panel technical report by \cite{robertson2019report} attributed this to a combination of internal creep and the loss of suction induced by a period of heavy rainfall from about October 2018 to the time of the failure. The catastrophic failure of Dam I caused a mudlfow of approximately 11 million $m^{3}$ of mining waste to flow rapidly downstream, destroying properties, infrastructure and agricultural land and entering the Paraopeba River \citep{f2020deformations}. Within days of the collapse, the mud covered an area of 3.13 x $10^{6}$ $m^{3}$ \citep{rotta20202019}, ultimately claiming the more than 250 lives and affecting the whole region's ecosystem \citep{porsani2019gpr}.
	
	The data used to analyse the precursory deformation and risk in this study comprises Sentinel-1 InSAR displacement time-series data for two overlapping descending orbit tracks (tracks 53 and 155), for the 17 months preceding the dam collapse. The relative line-of-sight displacement (RLOSD) time-series data covering the extent of Dam I at 20-m spatial resolution were generated using the Intermittent Small Baseline Subset (ISBAS) differential InSAR technique \citep{sowter2013dinsar} and made publicly available by \cite{grebby2021advanced}. Representative displacement time-series for areas exhibiting distinctive deformation within each of the main structural elements of the dam (i.e., dam wall, tailings beach) were obtained by extracting the time-series of a subset of contiguous pixels at each locality \citep{grebby2021advanced}. Figure~\ref{fig:studyA}b) shows the locations of the six areas (labelled locations 1-6) identified as exhibiting distinctive deformation over the dam. The number of contiguous pixels for which the time-series were extracted at locations 1 through to 6 is $30,~ 44,~ 46,~ 25,~ 38\mbox{ and }37$, respectively. 
	
	The time-series for locations 1, 2 and 3 cover the time period (sampling window) from 19 August 2017 until 17 January 2019 (track 155), while those for locations 4, 5 and 6 covered 24 August 2017 until 22 January 2019 (track 53). In both cases, displacement measurements were derived at a 12-day interval (sampling frequency). Accordingly, the multiple individual displacement time-series and the average time-series at each of the six locations were used to construct the machine-based algorithm for characterising and precursory monitoring of the tailings dam failure.

	%%%%%%%%%%%%%%%%%%%%%%%%%%%%%%%%%%%%%%%%%%
	
	\section{Methods}\label{sec:mth}
	The method can be summarised in the following three stages:
	(i) Monitoring of the RLOSD to identify the time of regime change;	(ii) Estimating state-of-the-system at the time of regime change; (iii) Sequential risk estimation and hazard classification. All the above steps build on monitoring the second order statistical properties of a time-series signal. These steps and the related concepts are described in the following subsections.
	
	\subsection{Statistical preliminaries}
	Conventional statistical modelling of a time-series, $\{Y_t\}$, often involves modelling its second order statistical properties, its trend -- slowly varying mean effect ($\mu_t$) -- the autocovariance ($\cov(Y_t, Y_{t+u}$)), and variance ($\sigma^{2}_t$), which are defined as:
	\begin{equation}\label{eq:secmom}
		\begin{aligned}
			\E(Y_t) & = \mu_t \\
			\var(Y_t) & = \E(Y_t -\mu_t)^2 = \sigma^2_t \\
			\cov(Y_t, Y_{t+u}) & = \E\{(Y_{t+u} -\mu_t)(Y_t -\mu_t)\}= c(t,t+u)
		\end{aligned}
	\end{equation}
	Key to such modelling is the notion of second order stationarity.
	
	\subsubsection{Second order stationarity}
	Let $\{Y_t\}$ be a time-series that is discretely sampled at $n$ regular time intervals $\{ t= 1,2, \hdots,n\}$, such that its population mean, variance ($\var(.)$) and auto-covariance ($\cov(.)$) are defined as in equation \eqref{eq:secmom}. Then, $Y_t$ is second order stationary if:
	
	\begin{equation}\label{eq:station}
		\begin{aligned}
			\E(Y_t) & = \mu \mbox{ (a constant)}\\
			\var(Y_t) & = \sigma^2 \mbox{ (a constant)}\\
			\cov(Y_t) & = E(Y_{t+u} - \mu)(Y_{t} - \mu) = c(u),\\ 
			&\mbox{ which are a function of time interval, $(t, t+u)$}. 
		\end{aligned}
	\end{equation}
A second order stationary time-series is a characteristic property that can be observed in the Fourier domain.	

	\subsubsection{Spectral density and periodogram}\
	If $Y_t$ is a second-order stationary time-series with the auto-covariance function, $c(u)$, satisfying $\sum_{u} \! |u|c(|u|) <\infty$, then $Y_t$
	has a continuous spectrum $f(\omega)$ that is defined as:
	
	\begin{equation}\label{eq:specdef}
		\begin{aligned}
			f(\omega) &= \sum_{u} \!e^{-i \omega u} c(u), -1/2 < \omega < 1/2, &\mbox{ admitting the inversion:}\\
			c(u)& = \int_{-1/2}^{1/2}\!e^{i \omega u} f(\omega) d(\omega).
		\end{aligned}
	\end{equation}
	
	The spectrum $f(\omega)$ is a unique signature of a second order stationary time-series marked by invariance over time. It partitions the total variance (power) of the time-series among sinusoidal event frequencies. Physicist Arthur Schuster \citep{schuster1906ii} proposed the periodogram to estimate the spectrum of a second order stationary time-series. The discrete Fourier transform (DFT) and periodogram of observed time-series data $y_t, t=1,2,\hdots,n$ are defined, respectively, by $d(\omega_{k}$) and $I(\omega_{k}$) as:
	
	\begin{equation}\label{eq:period1}
		\begin{aligned}
			d(\omega_{k}) &= \frac{1}{\sqrt{n}}\sum_{t=1}^{n}\!e^{-i\omega_{k} t} y_t,\\
			I(\omega_{k}) & = \frac{1}{{n}} \sum_{t=1}^{n}\!e^{-i\omega_{k} t} \hat{c}(u),\\
			\hat{c}(u)& = \frac{1}{n}\sum_{t=1}^{n-|u|}\!\{(y_t - \bar{y})(y_{t+|u|} - \bar{y})\}, \mbox{ where}\\
			& \omega_{k} = 2\pi k/n, k =0,1,2\hdots,n-1.
		\end{aligned}
	\end{equation}
	
	\subsection{Sequential quantification of non-stationarity}\label{seqquant}
	A non-stationary time-series is characterized by a spectrum that varies over time. The empirical methods for monitoring non-stationarity are described below.
	\subsubsection{Moving window mean and variance}\label{sec:movwn}
	Exploratory investigations into deviations from second order stationarity of a time-series $Y_t$ can be performed by estimating local sample means and variances of the time-series
	within subsets of time windows. If partitioning the length of the time-series $n$ into $L$ overlapping time windows $w_{t_{l}}, l =1,2,\hdots,L$ of equal length, $n_L$, the local 
	sample mean and variances of $Y_t$ in window $w_l$ are defined as:
	
	\begin{equation}\label{eq:sammv}
		\begin{aligned}
			\hat{m}_{w_{l}} &= \frac{1}{n_L}\sum_{t: t\in w_{l}}y_{t}
			\hat{v}_{w_{l}} &= \frac{1}{n_L - 1}\sum_{t: t\in w_{l}}\{y_{t} - 	\hat{m}_{w_{l}}\}^2.	
		\end{aligned}
	\end{equation}
	
	\subsubsection{Dynamic periodogram: evolving features of a time-series }\label{sec:dynper}
	$I_{s}(\omega_{k},l)$ is defined as the local periodogram (see equation \eqref{eq:period1}) of the time-series $Y_t$ partitioned in each of the $w_{t_{l}}, l: l=1,2...,L$ overlapping time windows of length $n_L$, for all sampling locations, $\{s: s=1,2,...,v\}$, across the Fourier frequencies, $\omega_{k} = 2\pi k/n_{L},\mbox{ } k = 1,2,...,n_{L}$. Accordingly, $\mathbf{I}_{s}$ denotes the complete periodogram matrix at any arbitrary location, $s$, defined as:
	
	\begin{equation}\label{eq:periodmat}
		\mathbf{I}_{s} = \left[
		\begin{array}{cccccc}
			I_{s}(\omega_{1},1)	& I_{s}(\omega_{1},2) & . & . & . & I_{s}(\omega_{1},L) \\
			I_{s}(\omega_{2},1)	& I_{s}(\omega_{2},2) & . & . & . & I_{s}(\omega_{2},L) \\
			.&  .&  .&  .&  .&  \\
			.&  .&  .&  .&  .&  \\
			.&  .&  .&  .&  .&  \\
			I_{s}(\omega_{n_{L}},1)	& I_{s}(\omega_{n_{L}},2) & . & . & . & I_{s}(\omega_{n_{L}},L)
		\end{array}\right]
	\end{equation}
	
	where, for simplicity, the local time window $I_{s}(\omega_{1},w_{t_{l}})$ is denoted by $I_{s}(\omega_{k},l)$. 
	Reordering the periodogram matrix in equation \eqref{eq:periodmat} for the periodograms at time window $w_{t_{l}}$ across all sampling locations $\{s: s=1,2,...,v\}$: 
	
	\begin{equation}\label{eq:periodmat2}
		\mathbf{I}_{w_{t_{l}}} = \left[
		\begin{array}{cccccc}
			I_{w_{t_{l}},s_1}(\omega_{1})	& I_{w_{t_{l}},s_1}(\omega_{2}) & . & . & . & I_{w_{t_{l}},s_1}(\omega_{n_{L}}) \\
			I_{w_{t_{l}},s_2}(\omega_{1})	& I_{w_{t_{l}},s_2}(\omega_{2}) & . & . & . & I_{w_{t_{l}},s_2}(\omega_{n_{L}}) \\
			.&  .&  .&  .&  .&  \\
			.&  .&  .&  .&  .&  \\
			.&  .&  .&  .&  .&  \\
			I_{w_{t_{l}},s_\nu}(\omega_{1})	& I_{w_{t_{l}},s_\nu}(\omega_{2}) & . & . & . & I_{w_{t_{l}},s_\nu}(\omega_{n_{L}})
		\end{array}\right]
	\end{equation}
	
	For a second order stationary time-series, $\mathbf{I}_{w_{t_{l}}}$ defines the signature of multiple time-series, $\mathbf{Y}_{t} = \{Y_{s_{1}}(t),Y_{s_{2}}(t),\hdots,Y_{s_{v}}(t)\}^{t}$ in time window $w_{t_{l}}$. If $\mathbf{Y}_{t}$ is composed of \emph{all} second order stationary time-series, $\mathbf{I}_{l_{1}}$ would be time invariant. However, statistically evolving time-series would lead to a time varying periodogram matrix, $\mathbf{I}_{l_{1}}$.
	
	\subsubsection{Feature space: PCA periodograms}\label{sec:pcasp}
	
	Principal components analysis (PCA) was introduced by Karl Pearson and has since become a standard unsupervised statistical learning method \citep{hastie2009elements}. A common use of PCA is for reducing the dimensionality of correlated primary features in multivariable dataset. If a dataset consists of $n$ --- potentially correlated --- features, the application of PCA leads to a smaller number $(m < n)$ of uncorrelated secondary features $\{Z_1,Z_2,…,Z_m\}$ that comprise linear combinations of the primary ($n$) features.
	
	It is well known that periodogram ordinates $I(\omega_k)$ of second order stationary time-series are asymptotically uncorrelated at Fourier frequencies, $\omega_k = 2 \pi k/n, k =0,1,2,...,n-1$ \citep{shumway2000time}. However, a non-stationary time-series has correlated periodograms. Here, PCA is used for two purposes: 
	\begin{itemize}
	 \item to reduce the dimension of periodogram vectors (i.e., spectral features in each time window, $w_l$) retaining most of the variation in the signal; 
	 \item to obtain uncorrelated secondary spectral features (see columns of equation \eqref{eq:periodmat2}, $\mathbf{I}_{w_{t_{l}}}$) for multiple RLOSD time-series.
	 \end{itemize}
	 Mathematically, this is achieved by solving the following optimization problem using a linear combination of periodogram columns of $\mathbf{I}_{w_{t_{l}}}$: 
	\begin{equation}\label{eq:pcaspec}
		\min A_{m} \var(Z_{s_{j},m}\mbox{ where }Z_{s_{j},m}= \sum_{i =1}^{n_l} \alpha_i I_{w_{t_{l}},s_1}(\omega_{i},\mbox{ such that }\sum_{i =1}^{n_l}\! \alpha^{2}_i =1
	\end{equation}
	Subsequently, the obtained uncorrelated secondary features are used to characterize the evolution of the state-of-the-system, using multivariate statistical learning methods, cluster analysis and classification.
	
	\subsubsection{State-of-the-system at regime change}\label{sec:clusan}
	The first stage in the stability monitoring process is to determine a time window within which the statistical features of the ground displacement signal undergo a structural change from a period of second order stationarity -- known as the time of regime change ($w_{t_{0}}$). A variety of statistical methods can be used to detect the transition of the system from a second order stationary to a non-stationary regime (e.g., frequentist, Bayesian). However, owing to the relatively short time-series in this retrospective analysis, an alternative approach is adopted here. Defining:
	\begin{equation}\label{eq:locvar}
	Var(w_{t_{l}}) = \sum_{i = 0}^{n_l/2}\!I_{w_{t_{l}},s_1}(\omega_{i}), l = 1,2,\dots,n.
	\end{equation}
	as the local variance of a time-series in window, $w_{t_{l}}$. Under second order stationarity, $Var(w_{t_{l}})$ remains approximately the same in all windows. However, a substantial departure indicates a change in the statistical regime, which can be identified by plotting $Var(w_{t_{l}})$ -- for the average RLOSD time-series for each of the six locations on the dam --  against time windows, $w_{t_{l}}, l = 1,2,\dots,L$.
	
	Inflection points on the the $Var(w_{t_{l}})$ plot are chosen as potential candidates for the time of regime change. From a pool of candidate inflection points, the most likely candidate for the time window of regime change ($w_{t_{0}}$) is determined using the principle of maximum inter-cluster variance \citep{das2019near}. To achieve this, the local periodogram in the window of regime change, $\mathbf{I}_{w_{t_{0}}}$, for all locations is estimated, as shown in equation \eqref{eq:periodmat2}, before using PCA to obtain a matrix of secondary spectral features, $\mathbf{Z}_{w_{t_{0}}}$, for each location $s=,1,2,\dots,v$. The feature matrix $\mathbf{Z}_{w_{t_{0}}}$ encapsulates the signature (i.e., slope stability) of the state-of-the-system (i.e., the dam) at $w_{t_{0}}$, and is derived from the average RLOSD time-series of all pixels at each location. This feature matrix is then used to partition the (six) sampling locations of the dam into a finite number ($k$) of clusters, $\mathbb{C} = \{C_1,C_2,\hdots,C_k\}$. $\mathbb{C}$ is assumed to be the \emph{baseline} stable state-of-the-system immediately preceding a dynamic transition into an unstable deformation regime. In the final stage, the evolution of the system is measured using a statistic that sequentially compares all subsequent spectral features $\mathbf{Z}_{w_{t_{0}}}$ against the baseline, $\mathbb{C}$.
	
	\subsection{Classification and risk}\label{sec:classrisk}
	In this final phase we derive risk thresholds based on a sequential classification methodology against the baseline feature matrix $\mathbb{C}$.
	\subsubsection{Classification}\mbox{}\\
	The following statistical law underpinning classification error was a key observation by \cite{das2019near}. Consider a classification problem where at any given point in time, $t$, it is desired to classify a group of observations $\{1,2,\dots,n\}$ based on features $\{I_1, I_2,\hdots,I_n\}$ into a finite number of classes, $\mathbb{C} = \{C_1, C_2,\hdots,C_k\}.$ Then:
	\begin{equation}\label{eq:lclassp}
	   \begin{aligned}
	    & p^{tj}_{Il} = Prob.(j \in C_l|I_j), l = 1,2,\dots,k; j =1,2,\dots,n\mbox{ such that}\\
	    & \sum_{l =1}^{k}\!p^{tj}_{Il} = 1.
	    \end{aligned}
	\end{equation}
	which denotes the probability of classification of a randomly selected observation $j =1,2,\hdots,n$ into class $C_{l}$ at time $t$. For simplicity, the superscript, $t$, is ignored henceforth.
	
	\subsubsection{Risk trajectory}
\begin{figure}[!h]
		\centering
		\includegraphics[height=80mm,width=90mm]{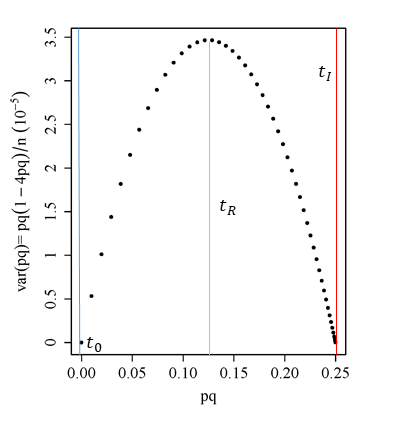}
		\caption{Relationship between the average classification uncertainty and its variance. The three vertical lines correspond to the three risk thresholds in time; $w_{t_{0}}$: time of regime change, $t_R$: time of emergent risk, and $t_I$: time of imminent risk.}
		\label{fig:parbol}
	\end{figure}

\mbox{}\\
		The classifier allocates the $j^{th}$ observation to class $C_z; z \in \{1,2,\dots,k\}$ if the class posterior probability $p_j$, given features $I_j$, satisfies:
	\begin{equation}\label{eq:clasprob}
	\begin{aligned}
	& p_j = \max_{l=1}^{k} p^{j}_{Il} = \mbox{ Prob}(I_{j} \in C_z),\mbox{ and}\\ 
	& q_j=1-p_j,\mbox{ that is, }q_j =\mbox{ Prob}(I_{j} \notin C_z)\\
	& z = arg \max_{l} p_{Il}, z \in \{1,2,\dots,k\},\mbox{ the estimated class}.
	\end{aligned}
	\end{equation}
	Defining $M(p_j) = p_jq_j, j =1,2,\dots,n$ to be the theoretical classification variation and $M(\hat{p}_j)$ as the corresponding maximum likelihood estimator, then \cite{das2019near} showed that expectation and variance of $\hat{p}_j$ share a parabolic relationship: 
	\begin{equation}\label{eq:rspro}
	 \begin{aligned}
	\E\{M(\hat{p}_j\} &= p_jq_j,\mbox{ and its variance},\\
	\var\{M(\hat{p}_j)\} &= \frac{1}{n}\{p_jq_j(1-4p_jq_j)\}
	\end{aligned}
	\end{equation}
	as shown in Figure~\ref{fig:parbol}.	
	The algorithm can then estimate two slope stability warning thresholds -- $t_R$ and $t_I$ -- based on the progression of this trajectory (Figure~\ref{fig:parbol}). The first milestone, $t_R$, is the time of an emergent instability risk, which corresponds to the theoretical maximum of the trajectory at $p_jq_j=0.125$, indicating a time of monotonic progression towards failure. This is followed by $t_I$, the time of imminent failure. The risk warning $t_I$ is estimated as the time window when the classification variance reaches close to its theoretical maximum that is, $p_jq_j=0.25$. On this basis, retrospective empirical estimates of $t_R$ and $t_I$ for the Brumadinho tailings dam collapse are generated. 
	
	Note that estimates of population parameters (e.g. $p$) based on $n$ samples $p_1, p_2,\dots, p_n$ are formally denoted by $\hat{p}$. Accordingly, functions of parameters such as $M(p)$ are formally denoted by $M(\hat{p})$. However, for brevity we drop \emph{hat} notation from the estimates in the remainder of the article.
	
	\subsection{Algorithm implementation}\label{sec:algorithm}
	Algorithm~\ref{alg:algotab} summarises the methodology for generating warnings of an emergent risk of failure ($t_R$) and an imminent risk of failure ($t_I$). A primary difference of this new algorithm compared with \cite{das2019near} is in the features used in monitoring. While \cite{das2019near} focused on first order properties of time-series, here the framework has is extended to include serially correlated and non-stationary time-series. The algorithm is implemented as a two-phase process using the statistical software \texttt{R} \citep{rprogram}. For visualization, the package \texttt{ggplot2} \citep{ggplot2} is used.
	
	\begin{algorithm}[!h]
		\SetAlgoLined
		\KwResult{\textbf{Phase I:} Feature estimating state of the system}
		\eIf{$w_t$ < $w_{t_{0}}$}{
			Estimate local periodogram, $I(\omega_{k},w_t)$\;
			Estimate a non-stationarity metric $S(w_t)$ based on $I(\omega_{k},w_t)$\;
			Repeat if $S(w_t)$ is not an inflection point\;
		}
		{   $w_t = w_{t_{0}}$, a candidate window of regime change\;
			Use PCA to construct secondary features from \;
			Cluster all locations at baseline $w_{t_{0}}$ into $\mathbb{C}$ clusters \;
			\eIf{Inter-cluster variance $\geq 80\%$ }
			{Move to classification phase}
			{Return to feature estimation}
		}
		\KwResult{\textbf{Phase II}: Classification and risk}
		At $j = t_{0+1}$ classify all locations into $\mathbb{C}$ class labels\;
		$p_l(t) = \mbox{class posterior probability, }q_l(t) = 1-p_l(t)$\;
		Define, $U_j = \mbox{median}_{l}p_l(j)q_l(j)\mbox{ classification variance}$\;
		\eIf{$U_j < 0.25$}
		{Repeat classification of locations into one of the baseline clusters for windows, $w_{j}, j = t_{0+1}, t_{0+2},\hdots,t_{0+m}$ \;
		}{
			Declare that slope failure is imminent
		}
		\caption{Algorithm for slope stability risk monitoring.}\label{alg:algotab}
	\end{algorithm}
	
	\section{Results and Discussion}\label{sec:res}
	\subsection{Estimating the time of regime change}\label{sec:ResTime}
\begin{figure}[!h]	
		%\widefigure
		%      \begin{subfigure}
		\centering
		\includegraphics[scale = 0.7]{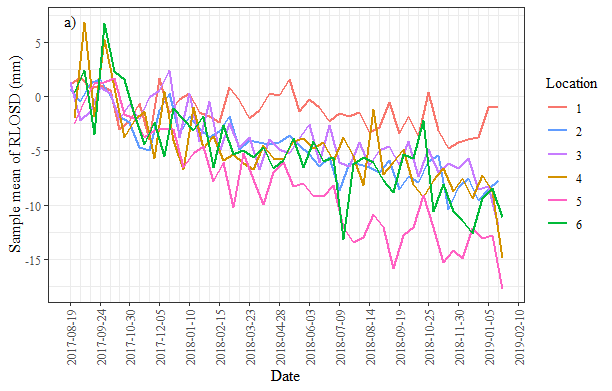} 
		%		\end{subfigure}\hfill
		\\ \vfill
		%		\begin{subfigure}
		% 		\centering
		\includegraphics[scale = 0.5]{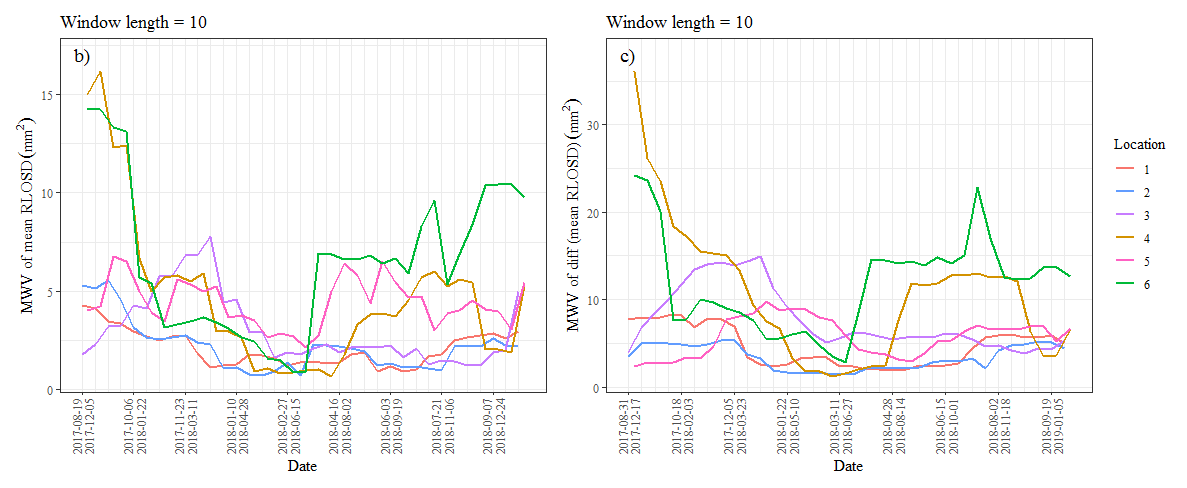} 
		%		\end{subfigure}\quad  
		\caption{Spatial trend and moving window variances at six sampling locations. \textbf{a}) Time-series of spatial sample mean of relative LOS displacement (RLOSD). \textbf{b}) Moving window spatial sample variance (MWV) of mean RLOSD. \textbf{c}) Moving window spatial sample variance (MWV) of the detrended mean RLOSD.}
		\label{fig:movwn}
	\end{figure}

	 The first phase of the algorithm is the estimation of the time of regime change, $w_{t_0}$, as described in section~\ref{sec:clusan}. Figures~\ref{fig:movwn}a) and b) show the trend and moving window variance of the time-series of the mean RLOSD observed at each location of the six locations, respectively. As previously outlined, the mean RLOSD was computed as the spatial mean of all the individual time-series extracted for the subsets of contiguous pixels at the six locations (see Figure~\ref{fig:studyA}), as defined by \cite{grebby2021advanced}. With the exception of location 1, all locations exhibit a monotonically decreasing (subsiding) displacement trend (Figure~\ref{fig:movwn}a). Furthermore, for locations 2--6, it is also apparent that both mean displacement and the moving window variances vary over time, which indicates non-stationary in second order statistical properties.

	% %\end{paracol}
	
	% Example of a figure that spans the whole page width (the commands %\widefigure and %\begin{paracol}{2}, %\linenumbers, and%\switchcolumn need to be present). The same concept works for tables, too.
	% %\begin{paracol}{2}
	%\linenumbers
	%\switchcolumn
	
	A key time-varying second order property is that of sample variance. An examination of Figure~\ref{fig:movwn}b) shows that the Brumadinho tailings dam underwent a deformation phase transition. Prior to 2 July 2018, the trajectory of spatial variance decreased in magnitude at each location, while after this date, increases in variance can be seen. The trend-adjusted moving window variances shown in Figure~\ref{fig:movwn}c) retain this property. This pattern is consistent across all six locations, albeit in varying magnitude. 
	
\begin{figure}[!h]	
		%\widefigure
		\centering
		\includegraphics[scale = 0.52]{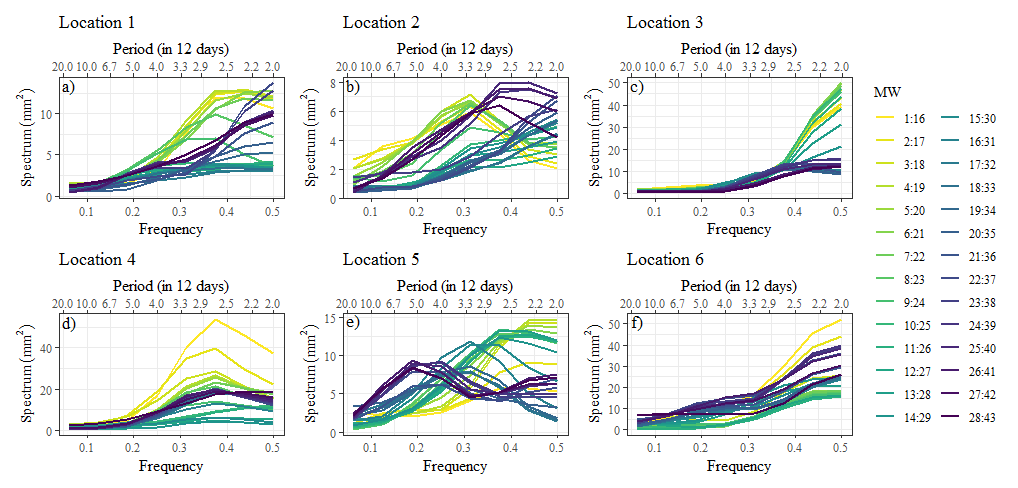} %image file here
		\caption{Local periodogram (window length of 16) of detrended mean RLOSD in successive time windows for each location.}\label{fig:fig5}
	\end{figure}

	As discussed in Section~\ref{sec:mth}, the temporal invariance of the spectral density function (equation \eqref{eq:specdef}) is the unique signature of a secondary order stationary time-series, whereas the periodogram (equation \eqref{eq:period1}) is the sample estimator of the spectrum. Thus, a common approach of assessing non-stationarity in a time-series is to estimate local periodograms in moving overlapping time windows and to then analyse the temporal variation of local periodograms. Periodograms of the mean adjusted RLOSD in local windows of several different lengths $(n_L)$ were estimated. In each local window $w_{t_{l}}, l = 1,2,\hdots,L$ we add one time unit (12 days) and remove the first time point of the previous window. The resulting local periodograms for each location for a local window length of 16 and overlaps of 15 time units are shown in Figure~\ref{fig:fig5}, where the $x-axis$ on the plots represents the Fourier frequencies, $\omega_{k}$. For a time-series of length $16$, it is only obtain possibly to obtain distinct estimates of spectra at $8$ Fourier frequencies due to aliasing \citep{nason2017should}. Figure~\ref{fig:fig5} clearly shows clear temporal variation in the local periodogram curves for each location. The evolving nature of the time-series is evident in that the periodogram ordinates appear more similar in some time windows -- in terms of both intercept and slope -- than in other windows. It is the $8$ periodogram ordinates that are used as features in the sequential slope stability monitoring algorithm.
	
	% %\end{paracol}
	% %\begin{paracol}{2}
	%\linenumbers
	%\switchcolumn
	
\begin{figure}[!h]
		\centering
		\includegraphics[scale = 0.55]{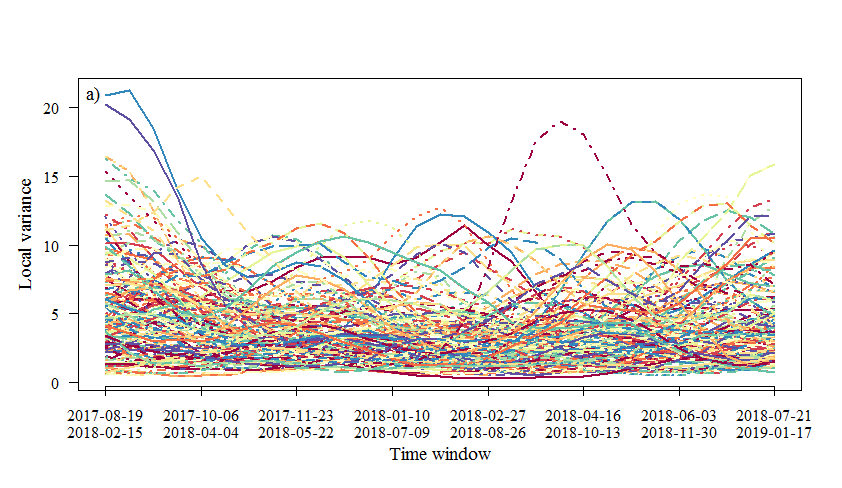} %image file here
		\includegraphics[scale = 0.55]{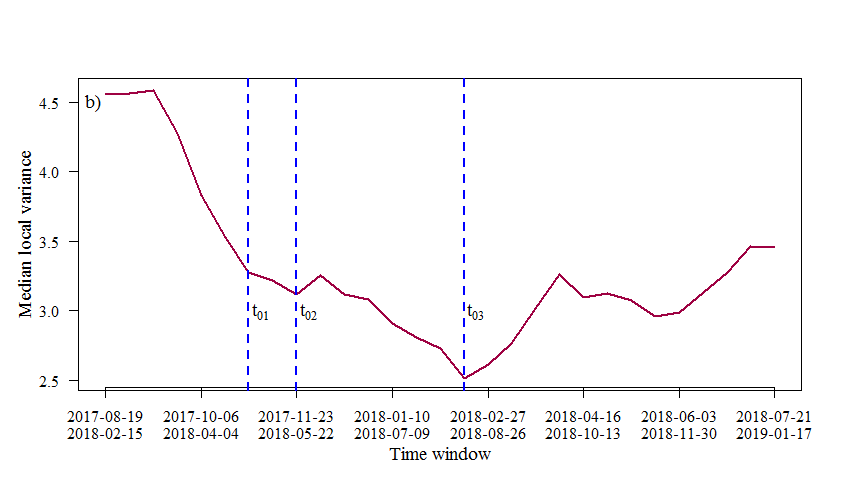} %image file here
		\caption{Local variance as sum of periodogram ordinates. \textbf{a}) Pixel-wise local variance of mean RLOSD at successive time windows. \textbf{b}) Median local variance of mean RLOSD at successive time windows.}\label{fig:fig6}
	\end{figure}

		The next key step in Phase I of the algorithm is to then utilise the periodograms to identify a time (window) of regime change -- i.e., a phase transition of the system from a stationary to non-stationary paradigm. To do this, \cite{das2019near} suggested implementing the non-stationarity metric on second order time domain properties proposed by \cite{das2016measuring} to assess deviation from stationarity. Here, however, a simpler approach that is applicable to spectral methods was adopted. It is well-known that the spectrum is a orthogonal partition of the variance of a time-series over Fourier frequencies \citep{shumway2000time}. Equivalently, the moving window spectra partitions the local variance -- within a particular time window -- among the Fourier frequencies. When a time-series is second order stationary, the spectrum is its unique signature and does not vary over time. This is observed as invariance in periodogram plots, over local time windows. In other words, the local variance -- obtained as the sum of local periodograms at Fourier frequencies -- must be uniform across all time windows. However, non-stationary time-series would have time varying (or evolving) local variance as shown in Figure~\ref{fig:fig6}a). 
	
	For all locations (1-6), the spatial median local variance across all contiguous pixels across moving time blocks were estimated to identify the points of inflection on the local variance trajectory (Figure~\ref{fig:fig6}b). Each point of inflection is considered to be a hypothetical point of regime change, $w_{t_{0}}$.  
	Accordingly, as shown in Figure~\ref{fig:fig6}b), three prospective time windows of regime change were identified, corresponding to windows $7$ (2017-10-30 : 2018-04-28), $9$ (2017-11-23 : 2018-05-22), and $16$ (2018-02-15 : 2018-08-14). At each prospective $w_{t_{0}}$, medoid clustering \citep{hu2021novel} of all pixels into a finite number of clustering with periodograms as features was performed, as described in section~\ref{sec:clusan}. The time window that led to the highest proportion of explained inter-cluster variation (with a threshold of $\approx 80\%$) for the fewest number of cluster partitions was chosen as $w_{t_{0}}$ \citep{das2019near}. The corresponding variations for the different numbers of cluster partitions are given in Table~\ref{tab:clustab}. With inter-cluster variation of 81\% for 4 cluster partitions, $w_{t_{0}} = 16$ (2018-02-15 : 2018-08-14) was selected as the optimal time of regime change. 
	
	%\end{paracol}
	\begin{table}[!h]
		\caption{Explained intra- and inter-cluster variations for different numbers of cluster partitions, $2, 3, 4, 5$, at various hypothetical times of regime change, $w_{t_{0}} = 7$ (2017-10-30 : 2018-04-28), $w_{t_{0}} = 9$ (2017-11-23 : 2018-05-22),\mbox{ and } $w_{t_{0}} = 16$ (2018-02-15 : 2018-08-14).}
		%\widefigure
		\centering
		\footnotesize
		\begin{tabular}{cccccccc}
			\hline
			\textbf{Time Window}               & \textbf{No. of Clusters} & \multicolumn{5}{c}{\textbf{Within Cluster   Variation}} & \textbf{Inter-Cluster Variation}(\%) \\ \hline
			\multirow{4}{*}{7}  & 2               & 0.304   & 0.153   &         &         &        & 54\%                        \\ 
			& 3               & 0.117   & 0.071   & 0.113   &         &        & 70\%                        \\ 
			& 4               & 0.041   & 0.034   & 0.078   & 0.071   &        & 78\%                        \\ 
			& 5               & 0.033   & 0.027   & 0.041   & 0.014   & 0.062  & 82\%                        \\ \hline
			\multirow{4}{*}{9}  & 2               & 0.182   & 0.245   &         &         &        & 57\%                        \\ 
			& 3               & 0.115   & 0.097   & 0.06    &         &        & 73\%                        \\ 
			& 4               & 0.012   & 0.075   & 0.058   & 0.038   &        & 82\%                        \\ 
			& 5               & 0.040    & 0.048   & 0.020    & 0.012   & 0.021  & 86\%                        \\ \hline
			\multirow{4}{*}{16} & 2               & 0.148   & 0.32    &         &         &        & 53\%                        \\  
			& 3               & 0.106   & 0.072   & 0.071   &         &        & 75\%                        \\ 
			& 4               & 0.018   & 0.042   & 0.06    & 0.072   &        & 81\%                        \\ 
			& 5               & 0.060    & 0.005   & 0.018   & 0.024   & 0.042  & 85\%                        \\ \hline
		\end{tabular}
		\label{tab:clustab}
	\end{table}  
	%\begin{paracol}{2}
	%\linenumbers
	%\switchcolumn

\subsection{Risk of failure warnings}\label{sec:ResPred}
	
\begin{figure}[!h]
		%\widefigure
		\centering
		\includegraphics[scale = 0.52]{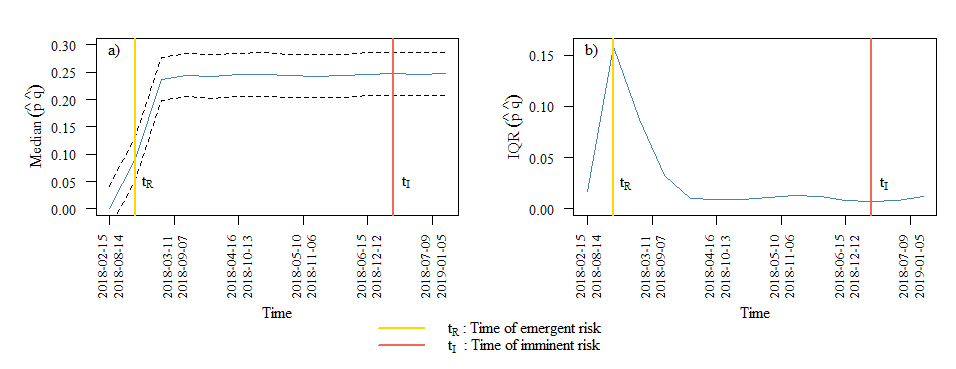} %image file here
		\caption{Risk time-series and risk thresholds $t_R$ (in gold) and $t_I$ (in red) for $w_{t_{0}} =16$ (2018-02-15 : 2018-08-14). \textbf{a}) Risk of failure, $Median(pq$). \textbf{b}) Variability of risk of failure, $IQR (pq)$.}\label{fig:fig7.1}
	\end{figure}

	Using local periodograms for a window length of $16$ as features, all time-series were subsequently classified into the class labels of $w_{t_{0}}$ at all subsequent points of time $w_{t_{0+1}}, w_{t_{0 +2}}, \hdots$. Mis-classification ($pq$) is treated as an indication of the evolution of the state-of-the-system. At each time window, the spatial median of mis-classification and the corresponding interquartile range (IQR) for $pq$ were calculated. The time-series of $Median(pq)$ and $IQR(pq)$ for $w_{t_{0}}=16$ (2018-02-15 : 2018-08-14) are shown in Figure~\ref{fig:fig7.1}. Based on the relationship between $pq$ and its variability, retrospective warnings for the time windows of emergent ($t_R$) and imminent risk of failure ($t_I$) for the Brumadinho tailings dam were 27 February 2018--26 August 2018 and 27 June 2018--24 December 2018, respectively. Consequently, the algorithm identifies a risk of imminent failure of the dam at least a month prior to the actual event. This is comparable with the 40 days of advanced warning based on the inverse velocity method by \cite{grebby2021advanced}. If monitored using this approach in near real-time, this would have provided adequate warning for detailed site investigations or risk mitigation measures to have been implemented. Moreover, the warning of an emergent risk is even earlier, at least five months prior to the collapse, coinciding with an initial phase of accelerating deformation highlighted by \cite{f2020deformations}. This is considerably earlier than the earliest reliable indication of an emerging failure 51 days prior based on the inverse velocity method \citep{grebby2021advanced}. Although, with hindsight, this deformation could correspond to a period of secondary creep rather than tertiary, at the time, such a warning would still have been useful in initiating crucial follow-up assessments of the stability of the dam. 
	
	%\end{paracol}
	%\begin{paracol}{2}
	%\linenumbers
	%\switchcolumn
\subsection{Effect of time windows on risk of failure warnings}
 \label{sec:reswin}
 \subsubsection{Local periodograms}
The discussion thus far has focused on the results associated with the estimation of local periodograms over time windows of length 16. In order to assess the effect of the length of the moving window on the regime change and risk estimation, several other window lengths were used in the computations. These included periodogram estimates for the same time-series over moving windows of length $24$ and $32$, with varying overlap lengths.

% Figure~\ref{fig:dynsp.24} shows heatmaps displaying local periodograms for overlapping time windows. Each time series is of length $24$. Overlap length is $18$ time units. For second order stationary time series we expect similar color intensity along the $-y-axis$ for a fixed value on the $x-axis$.  We replicate our computations with windows of length 32 with overlap length 31 shown using superimposed line plots in Figure~\ref{fig:locspec32}. 

\begin{figure}[!h]
		\centering
		\includegraphics[scale = 0.5]{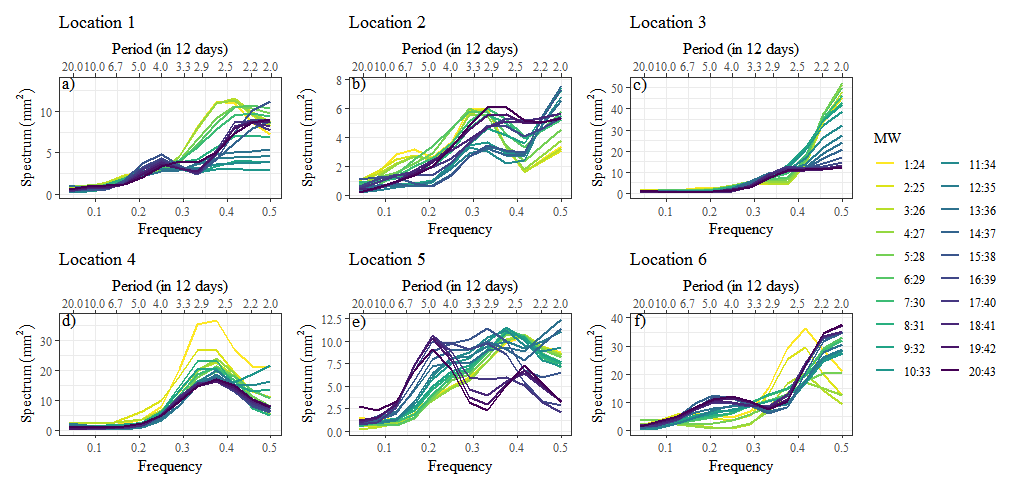} %image file here
		\caption{Local periodogram (window length of 24) of detrended mean RLOSD in successive time windows.}\label{fig:locspec24}
	\end{figure}

\begin{figure}[!h]
		\centering
		\includegraphics[scale = 0.5]{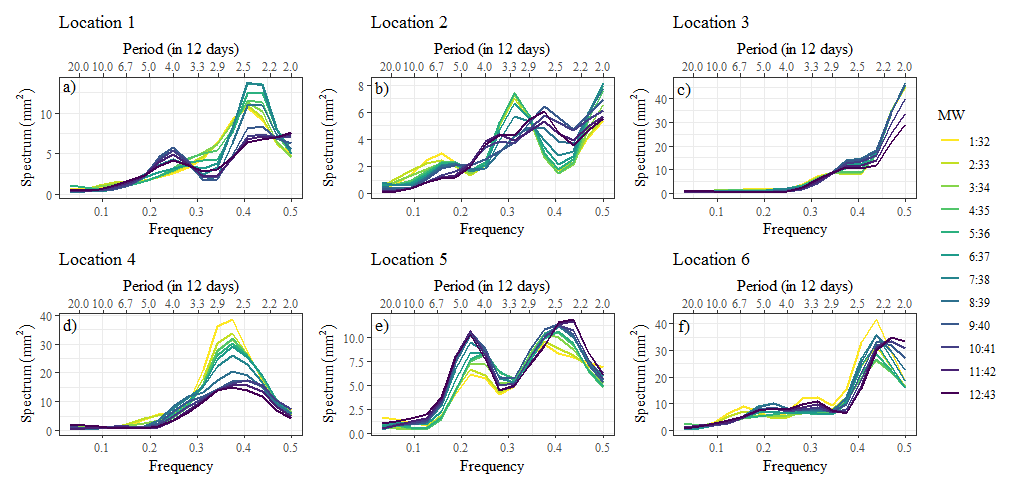} %image file here
		\caption{Local periodogram (window length of 32) of detrended mean RLOSD in successive time windows.}\label{fig:locspec32}
	\end{figure}

Comparing Figures~\ref{fig:locspec24} and \ref{fig:locspec32} with Figure~\ref{fig:fig5}, it can be observed that all three window estimates show temporal variation (and clustering) of local periodogram curves. The primary differences are two-fold:
\begin{itemize}
    \item longer windows allow for a broader spectrum of variation. That is, when choosing local periodograms for time-series of length $32$ it is possible to estimate (low frequency) periodic components at twice the length of that allowed by time window of length $16$ (1 time unit = 12 days).
    \item Accordingly, the local window (numbers) of clustering vary. 
\end{itemize}
Overall, the analysis reveals that all time window lengths estimated the maximum variance to coincide with the Fourier frequency that corresponds to the period of July--August 2018. This confirms that the window length over which the local periodograms are estimated have negligible effect on the prospective time windows of regime change and, hence, the subsequent timing of the risk of failure warnings. However, this may not necessarily always be true, and so a more objective window selection approach would be beneficial in generalising the algorithm for application to a broader range of cases.

The key to objectively optimising window selection is a methodology that can estimate periodograms at all frequencies that have significant variance components, for given a slope failure process. In other other words, such a method must allow estimation of power, inherent in the RLOSD time-series, at the smallest typical frequency or longest time period typical of that type of slope failure, conditional on the sampling frequency. Examples of window width estimation (as also a window kernel) include those from within the statistical sciences and signal processing (equivalent to \emph{bandwidth}), such as density estimation \citep{green1993nonparametric,bowman1984alternative}, trend estimation \citep{fried2004robust}, and periodogram estimation \citep{ombao2001simple}. As demonstrated by many of these examples, a rigorous and automatic approach can be developed using the resampling methodology known as cross-validation. One potential solution, based on mechanistic \emph{apriori} knowledge of prevalent Fourier spectra of the RLOSD time-series for a given type of slope failure, would be to use a cross-validation-based decision criteria to select the optimal window length by minimizing a loss function which will ensure that the spectra at the lowest relevant frequency can be estimated.

% 	\begin{figure}[!h]
% 		\centering
% 		\includegraphics[scale = 0.2]{Figures/Figure 5_Dynamics_spectra.png} %image file here
% 		\caption{Heatmaps of local periodograms (window size of 24 and overlap length of 18) of detrended mean RLOSD.}\label{fig:dynsp.24}
% 	\end{figure}
	
	\subsubsection{Time of regime change windows}\label{app:pq est}
\begin{figure}[!h]
		\centering
		\includegraphics[scale = 0.5]{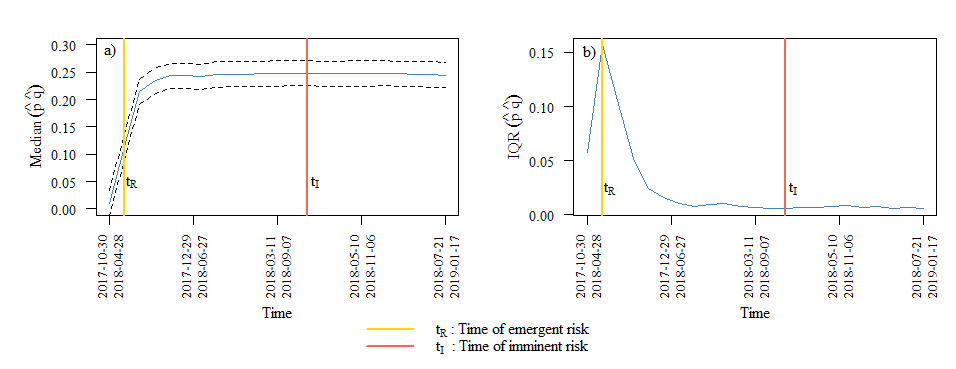} %image file here
		\caption{ Panels show the trajectories of risk statistic with time and risk thresholds $t_R$ (in gold) and $t_I$ (in red) for $w_{t_{0}} =7$ (2017-10-30 : 2018-04-28). \textbf{a}) Risk of failure, $Median(pq$). \textbf{b}) Variability of risk of failure, $IQR (pq$).}\label{fig:win7}
	\end{figure}

\begin{figure}[!h]
		\centering
		\includegraphics[scale = 0.5]{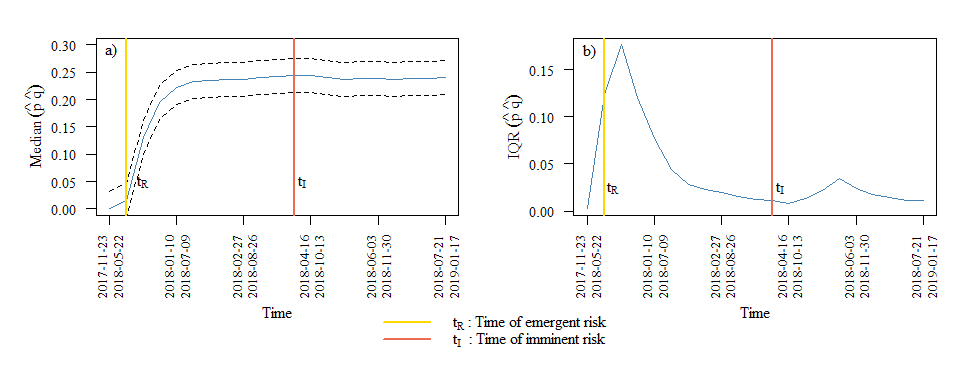}
		\caption{ Panels show the trajectories of risk statistic with time and risk thresholds $t_R$ (in gold) and $t_I$ (in red) for $w_{t_{0}} =9$ (2017-11-23 : 2018-05-22). \textbf{a}) Risk of failure, $Median(pq)$. \textbf{b}) Variability of risk of failure, $IQR (pq)$.}
		\label{fig:win9}
	\end{figure}

	As previously shown in Figure~\ref{fig:fig6}b), three inflection points were identified and then subjected to cluster analysis (Table~\ref{tab:clustab}) to estimate the optimal time window of regime change ($\hat{w}_{t_{0}}$). As a result, ${w}_{t_{0}} = 16$ (2018-02-15 : 2018-08-14) was selected. However, to assess the effect of the time window of regime change on the risk of failure warnings, estimates of the risk thresholds, $t_R$ and $t_I$, were computed for the two alternative prospective time windows of $w_{t_{0}} = 7$ (2017-10-30 : 2018-04-28) and $9$ (2017-11-23 : 2018-05-22). Figures~\ref{fig:win7} and \ref{fig:win9} show the time-series of the median and interquartile range (IQR) for the risk statistic ($pq$) and risk thresholds $t_R$ and $t_I$ for $w_{t_{0}} =7$ and $w_{t_{0}} =9$, respectively. As for $w_{t_{0}} = 16$, it can be seen that both alternative candidates for the time of regime change also predict an imminent risk of failure ($t_I$) well in advance of the actual collapse of the dam. For $w_{t_{0}} =7$, an imminent risk of failure is detected in the time window 11 March 2018--7 September 2018, while for $w_{t_{0}} = 9$ this risk is detected during the period 16 April 2018--13 October 2018. These risk warnings precede that for the optimal window for the time of regime change (${w}_{t_{0}} = 16$) by 2--3 months and the actual collapse by several months. However, as also observed by \cite{das2019near}, these earlier time windows of regime change ($w_{t_{0}} = 7$ and $w_{t_{0}} = 9$) produce maximum values for the risk statistic ($pq$) that are less than that for both $w_{t_{0}} = 16$ and the theoretical maximum of 0.25. Consequently, the risk warnings generated for the earlier alternative time windows are less accurate - that is, more prone to false positive - than those for the optimal time of regime change window of 16.
	
	Although in this case the cluster analysis was able to identify the optimal time of regime change window from three prospective candidates, a more robust approach could be more effective in addressing the issue of identifying a unique solution. The problem of detecting a time of regime change is similar in essence to that of detecting a structural change point in a time-series. With regards to change point detection, there are numerous potential approaches from both the frequentist and Bayesian perspectives that require future investigation \citep{killick2012optimal, cho2012multiscale, cho2015multiple, ghassempour2014clustering}. Alternately, detecting $\hat{w}_{t_{0}}$ can also be considered analogous to detecting anomalies in streaming data, as in \cite{hill2009real} and \cite{dereszynski2011spatiotemporal}.
	
	%%\end{paracol}
	\section{Conclusions}\label{sec:conclusion}
	Conventional approaches to predicting slope failures are based on the inverse velocity method. However, predictions made using this method can be unreliable for risk management owing to its highly subjective nature and the fact that the deterministic relationships of this exponential law often materialises close to the occurrence of a failure, therefore offering limited prior warning of a potential catastrophe. In this study, an alternative risk identification approach is proposed, which incorporates theories from spectral analysis of time-series to extend a data-driven sequential monitoring algorithm developed by \cite{das2019near}. This statistical-based algorithm is more objective and can be applied to any displacement time-series, sampled at regular time intervals, accounting for serial correlation.
	
	The efficacy of this new algorithm has been demonstrated through application to satellite InSAR displacement monitoring data associated with the 2019 Dam I Brumadinho tailings dam collapse. The algorithm analyses second order properties to show that the tailings dam transitioned into an unstable epoch around July 2018, several months before the actual collapse. This was evident in the moving window variance plots and further corroborated using the periodogram-based local variance plots; the latter of which accounts for serial correlation within each displacement time-series. Whereas most slope stability monitoring analysis only considers the trend or first order moments of the deformation signal, the results of this study shows that investigating second order statistical properties is critical. 
	
	The optimum time of regime change window, $w_{t_{0}}$ = $16$ (2018-02-15 : 2018-08-14), was identified based on cluster analysis and then subsequently used to retrospectively estimate risk of failure warnings for Dam I. The algorithm detected an emerging risk of failure during the period 27 February 2018--26 August 2018 and an imminent (maximum) risk of collapse of the tailings dam during 27 June 2018--24 December 2018. The latter warning comes at least several weeks prior to the actual dam failure on 25 January 2019, while the first sign of an emerging risk is evident at least five months prior to the failure. The amount of advanced warning for the risk of imminent failure provided by the algorithm is comparable to previous analysis based on the inverse velocity method, although considerably longer in terms of reliably detecting signs of an emerging failure. This is ultimately due to the statistical nature of the algorithm making it more objective and robust than empirical approaches, which also means that it can be readily applied to other slope failures. Although the risk warnings generated here are based on retrospective analysis, application of the algorithm on near real-time monitoring data could have provided sufficient early-warning to enable detailed site investigations or more urgent risk mitigation measures to have been implemented in order to advert a humanitarian and environmental catastrophe. Overall, the results of this study further attests the role that satellite InSAR can have in integrated dam monitoring and early warning systems.
	
	Further work is required to make the algorithm more widely applicable to analysing the risk associated with any potential slope failure scenario. This would primarily include the implementation of more objective or automated approaches for selecting the optimal time window length and time window of regime change. Additionally, the algorithm may also be further augmented by considering both spatial and temporal auto-correlation, in order to develop a spatio-temporal risk statistic. This would enable not just a temporal warning of a risk of failure, but also highlight the specific location of the instability.

	\vspace{6pt}
	
	%%%%%%%%%%%%%%%%%%%%%%%%%%%%%%%%%%%%%%%%%%
	%% optional
	%\supplementary{The following are available online at \linksupplementary{s1}, Figure S1: title, Table S1: title, Video S1: title.}
	
	
	%%%%%%%%%%%%%%%%%%%%%%%%%%%%%%%%%%%%%%%%%%
	
%	\acknowledgments{The authors are grateful to Dr. Stephen Grebby for kindly sharing the processed InSAR RLOSD data.}
	
%	\conflictsofinterest{The authors declare no conflict of interest.} 
	
%	\reftitle{References}
	
	
%	\externalbibliography{yes}
	\bibliography{Manuscript_Brazil_wip_home.bib}

\end{document}